\documentclass[a4paper,20pt]{article}
    \usepackage[T1]{fontenc}
    \usepackage{graphicx}
    \usepackage{epsfig}
    \usepackage{amsmath,marvosym}
    \usepackage{amsfonts}
    \renewcommand{\abstract}{}
\begin{document}
\makeatletter
\renewcommand{\@oddhead}{\textit{YSC'17 Proceedings of Contributed Papers} \hfil \textit{A. Simon, A. V. Bondar, V. M. Reshetnyk}}
\renewcommand{\@evenfoot}{\hfil \thepage \hfil}
\renewcommand{\@oddfoot}{\hfil \thepage \hfil}
\fontsize{11}{11} \selectfont

\title{The first spectra for the RX~J0440.9+4431 from 2m Terskol telescope}
\author{\textsl{A. Simon$^{1}$, A. V. Bondar$^{2}$}, V. M. Reshetnyk$^{1}$}
\date{}
\maketitle
\begin{center} {\small $^{1}$National Taras Shevchenko University of Kyiv, Glushkova ave., 4, 03127, Kyiv, Ukraine\\
$^{2}$International Center for Astronomical, Medical and Ecological Research Terskol Observatory\\
andrew\_simon@mail.ru}
\end{center}

\begin{abstract}
We present the first results on the spectra of Be/X-ray binary RX
J0440.9+4431 obtained with the 2m Ritchey-Cretein-Coude telescope
with Cassegrain Multi Mode Spectrograph (CMMS) (with $R=14000$) at
Terskol observatory. The H$_{\alpha}$ line profile indicates that
the new episode of the V/R variability is occuring in the system.
The profiles of the H$_{\alpha}$, H$_{\beta}$ and HeI 7065.71,
6678.15, 5875.97 lines were analyzed and equivalent width were
determined. We compared our H$_{\alpha}$ line profile parameters
with the previous results from the literature \cite{Reig05} and
estimated characteristic time scale for disc changes as about
14~years.
\end{abstract}

\section*{Introduction}

\indent \indent RX J0440.9+4431 belongs to the most numerous class
of high-mass X-ray binaries --- Be/X-ray binaries. The system has
two components: a blue main sequence star B0.2Ve \cite{Reig05} and
neutron star orbiting around it. This object belongs to the
relatively bright X-ray sources with $V=10.78$. Its spectra show a
long-term variability of the Balmer and HeI double-peaked emission
lines. As an X-ray pulsar it has a pulsing period of 202~s
\cite{Reig99} and suspected orbital period over 150~days. The
distance to the object was estimated as about 3.3~kpc \cite{Reig05}.

\section*{Observations}

\indent \indent Spectral observations were provided with the help of
the 2m Ritchey-Cretein-Coude telescope with CMMS at Terskol
observatory on February 2--3, 2010. Spectrograph was equipped with
the diffraction grating with 300 grooves per millimeter and a
blazing angle of 8$^{\circ}$. During these observations three
echelle spectra with 30-minutes exposures were obtained. From these
spectra a median spectrum was compiled for further analysis. This
spectrum contains 28 echelle orders and covers a spectral range of
$3908-7507$~{\AA}. Within this range the dispersion varies from
0.12~{\AA}/px to 0.25~{\AA}/px. But due to the low signal-to-noise
ratio in blue region we used for our analysis only the data with
$\lambda>5000$~{\AA}. In the region of the H$_{\alpha}$ line
signal-to-noise ratio is $\approx100$.

The data reduction was performed with the help of \texttt{Dech95}
software and spectra analysis was provided with \texttt{Dech20T}
software \cite{Gala92}, \cite{GalaWeb}.

\section*{Results}
\indent \indent We found and analyzed five double-peaked symmetric
class 1 (according to classification in \cite{Hanu95}) emission
lines in the spectrum: H$_{\alpha}$, H$_{\beta}$ and HeI 7065.71,
6678.15, 5875.97 lines. We should mention that the last asymmetric
profile for the H$_{\alpha}$ line was seen in 1996. For all of these
lines we calculated $\log(V/R)$, equivalent width, mean intensity of
the blue and red peaks over the central depression and peaks
separation. All these values are presented in Table~\ref{Andrew}. As
we can see, all our data for the H$_{\alpha}$ line are in good
agreement with \cite{Reig05}.

We also studied the radial velocities fields and general structure
of the line profiles. All of these lines are shifted blueward. First
of all we found that some of the V and R peaks have their own two
peaks --- right and left ones with a gap between them. Except this,
our attention was attracted to the features with width of about
94~km/s and 111~km/s in HeI lines: for these velocities the
intensity of HeI 7065.71 and HeI 5875.97 anti-correlate with the
intensity of HeI 6678.15, but it is necessary to confirm this fact
in the further investigations. It is worth to note that the same
lines have different sign for $\log(V/R)$ value (see
Table~\ref{Andrew}).

After analyzing the data from \cite{Reig05} (Table~1 and Fig.~3) and
comparing those data with our ones we can make some assumptions
about time scale of disc changes. EW and the shape of H$_{\alpha}$
line profile are very similar to the profiles at the end of 1995 and
beginning of 1996. In addition, H$_{\beta}$ profile in our data,
first time after the end of the 1997, shows emission with the
approximate value of EW$\approx-1.2$~{\AA} - the biggest negative
value for the whole history of the object observations (since March,
1, 1996 for H$_{\beta}$ line). Taking into account all these facts
we can suggest that the characteristic time scale of the disk
changes in RX~J0440.9+4431 system is about or more than 14~years.

\begin{table}[h]
\caption{\label{Andrew}Emission line parameters}
\begin{center}
\begin{tabular}{ccccc}
\hline

Line & $\log(V/R)$ & EW, \AA & $\bigtriangleup_{peak}$, km/s & $I_{p}/I_{cd}$\\
\hline
HeI 7065.71 &-0.12&-1.05$\pm$0.10&406&1.23\\
\hline
HeI 6678.15 &0.05&-0.65$\pm$0.10&284&1.10\\
\hline
HeI 5875.97 &-0.31&-0.78$\pm$0.10&342&1.14\\
\hline
H$_{\alpha}$ &-0.06&-9.8$\pm$0.8&298&1.20\\
\hline
H$_{\beta}$ &-0.21&-1.18$\pm$0.05&365&1.24\\
\hline
\end{tabular}
\end{center}
\end{table}

\section*{Conclusions}
\indent \indent The results of the preliminarily analysis of the
Be/X-ray binary RX~J0440.9+4431 spectra, observed at Terskol
observatory are in a good agreement with previous results by other
authors. The appearance of the asymmetric profiles of emission lines
indicates that the new epoch of V/R variability has been started.
This fact allows us to conclude that the time scale of the disc
evolution is about 14~years or larger.

\section*{Acknowledgement}
\indent \indent This work was supported by the International program
``Astronomy in the Elbrus region (2010-2014)'' of ICAMER. We thanks
the stuff of the ICAMER for the help in observations. V.~M.~Ivchenko
for the valuable discussions, E.~A.~Karitskaya and N.~G.~Bochkarev
for the help and support. Especial thanks to Ju.~Kuznetzova for the
help in the data reduction.

\end{document}